\documentclass[a4paper]{jpconf}

\usepackage{subfigure}
\usepackage{multicol}
\usepackage{amsmath}
\usepackage{epsfig}
\usepackage{graphicx}
\usepackage{color}
\usepackage{microtype}
\usepackage{hyperref}
\usepackage[dvipsnames]{xcolor}
\usepackage{framed}

\usepackage{amsmath,amsfonts,amssymb,mathrsfs,mathabx}
\usepackage{dsfont}
\usepackage{tikz}
\usetikzlibrary{positioning,shapes,arrows}
\hypersetup{
    unicode=true,             
    pdftoolbar=false,         
    pdfmenubar=false,         
    pdffitwindow=false,       
    pdfstartview={FitH},      
    pdftitle={},              
    pdfauthor={},             
    pdfsubject={},            
    pdfcreator={},            
    pdfproducer={},           
    pdfkeywords={},           
    pdfnewwindow=true,        
    colorlinks=true,          
    linkcolor=red,            
    citecolor=blue,           
    filecolor=green,          
    urlcolor=blue             
}

\newcommand{\To}{\longrightarrow}

\begin{document}
\title{\sf Solvable spectral problems from 2d CFT and ${\cal N}=2$ gauge theories\footnote{Based on 
a talk given at the XXVth International Conference on Integrable Systems and Quantum Symmetries, 
June 6--10, $2017$, Prague, Czech Republic; to be published in {\it Journal of Physics:~Conference
Series.}}}

\author{\sf M~R Pi\c{a}tek$^{1,3}$ and A~R Pietrykowski$^{2,3}$}

\address{$^1$ Institute of Physics, University of Szczecin, Wielkopolska 15, 70--451 Szczecin, Poland}
\address{$^2$ Institute of Theoretical Physics, University of Wroc{\l}aw, M.~Borna 9, 50--204 Wroc{\l}aw, Poland}
\address{$^3$ Bogoliubov Laboratory of Theoretical Physics, JINR, 141980 Dubna, Russia}

\ead{piatek@fermi.fiz.univ.szczecin.pl, radekpiatek@gmail.com, pietrie@theor.jinr.ru}

\begin{abstract}
The so-called 2d/4d correspondences connect two-dimensional conformal field theory (2d CFT), 
${\cal N}=2$ supersymmetric gauge theories and quantum integrable systems. 
The latter in the simplest case of the SU(2) gauge group are nothing but the quantum-mechanical systems. 
In the present article we summarize our recent results and list open problems concerning an 
application of the aforementioned dualities in the studies of 
spectral problems for some Schr\"{o}dinger operators with Mathieu--type periodic, 
periodic PT--symmetric and (Heun's) elliptic potentials. 
\end{abstract}

\section{Introduction}
They have long been known connections between ${\cal N}=2$ 
supersymmetric Yang--Mills theories and integrable models.
Let us consider perhaps the simplest example of such a relationship, 
i.e., a link between the SU(2) pure gauge Seiberg--Witten theory and 
the one-dimensional sine-Gordon model (cf.~\cite{Gorsky:1995zq}). 
Concretely, the statement here is that the Bohr--Sommerfeld periods 
$$
\Pi(\Gamma) \equiv \oint\limits_{\Gamma}P_{0}(\varphi)\,{\rm d}\varphi
=\oint\limits_{\Gamma}\sqrt{2(u-\hat\Lambda^2\cos\varphi)}\,{\rm d}\varphi
$$
for the classical sine-Gordon model defined by 
${\cal L}_{\rm sG}=\frac{1}{2}\dot{\varphi}^2-\hat\Lambda^2\cos\varphi$,
and for two complementary contours $\Gamma=A, B$ encircling two turning points 
$\pm\arccos(u/\hat\Lambda^2)$ define the Seiberg--Witten system
\cite{Seiberg:1994rs}:
$a=\Pi(A)$,
$\partial {\cal F}(a)/\partial a=\Pi(B)$.
Here, $a$ is a modulus and ${\cal F}(a)$ denotes 
the Seiberg--Witten prepotential determining the low energy effective 
dynamics of the four-dimensional ${\cal N}=2$ supersymmetric SU(2) pure gauge theory.

As has been observed in \cite{Mironov:Morozov:2010} (see also \cite{He:2010zzc,Maruyoshi:2010})
the above statement has its `quantum analogue' or `quantum extension' 
which can be formulated as follows. Namely, the monodromies ({\it exact} BS periods)
$$
\widetilde{\Pi}(\Gamma)\equiv\oint\limits_{\Gamma}P(\varphi, \hbar)\,{\rm d}\varphi,
\;\;\;\;
P(\varphi, \hbar)=P_{0}(\varphi)+P_{1}(\varphi)\hbar+P_{2}(\varphi)\hbar^2+\ldots
$$
of the exact WKB solution
\begin{equation}\label{M}
\psi(\varphi)=\exp\left\lbrace\frac{i}{\hbar}\int\limits^{\varphi}
P(\rho, \hbar)\,{\rm d}\rho\right\rbrace
\;\;\;\rm{to\;the\;eq.}\;\;\;\boxed{
\left[-\frac{\hbar^2}{2}\frac{\partial^2}{\partial\varphi^2}+\hat\Lambda^2\cos\varphi\right]\psi(\varphi)
\;=\;\mathrm{E}\,\psi(\varphi)}
\end{equation}
define the Nekrasov--Shatashvili system \cite{NS:2009}:
$a=\widetilde{\Pi}(A)$,
$\partial {\cal W}(\hat\Lambda, a,\hbar)/\partial a=\widetilde{\Pi}(B)$.\footnote{For 
SU(N) generalization of this result, see \cite{Mironov:2009dv}.}
Here, ${\cal W}(\hat\Lambda, a, \hbar)={\cal W}_{\rm pert}(\hat\Lambda, a, \hbar)
+{\cal W}_{\rm inst}(\hat\Lambda, a, \hbar)$
is the effective twisted superpotential of two-dimensional
SU(2) pure gauge ($\Omega$-deformed) SYM theory defined in \cite{NS:2009}
as the following Nekrasov--Shatashvili (NS) limit
$$
{\cal W}(\hat\Lambda, a, \hbar)\;\equiv\;
\lim_{\epsilon_2\to 0}\epsilon_2\log{\cal Z}(\hat\Lambda, a, \epsilon_{1}\!\!=\hbar,\epsilon_2)
$$
of the Nekrasov partition function
${\cal Z}(\hat\Lambda, a, \epsilon_1, \epsilon_2)={\cal Z}_{\rm pert}(\hat\Lambda, a, \epsilon_1, \epsilon_2)
{\cal Z}_{\rm inst}(\hat\Lambda, a, \epsilon_1, \epsilon_2)$ \cite{Nekrasov:Okounkov:2003,Nekrasov:2002qd}.

Interestingly, on the other hand the Mathieu equation written in (\ref{M}) 
is nothing but the Schr\"{o}dinger equation for the 
sine-Gordon model. One can show that the energy eigenvalue is determined 
by the SU(2) pure gauge twisted superpotential ($\hbar=\epsilon_{1}$) \cite{Mironov:Morozov:2010,PP1}:
\begin{eqnarray*}
2\textrm{E}_{\rm sG}=2\textrm{E}(a) &=&
a^2
+\hat\Lambda^4\left(
\frac{1}{2\, a^2}
+\frac{\epsilon _1^2}{8\, a^4}
+\frac{\epsilon _1^4}{32\, a^6}
+\frac{\epsilon _1^6}{128\, a^8}
\right)
\\
&+&\hat\Lambda^8 \left(
\frac{5}{32\, a^6}
+\frac{21\, \epsilon_1^2}{64\, a^8}
+\frac{219\, \epsilon_1^4}{512\, a^{10}}
+\frac{121\, \epsilon_1^6}{256\, a^{12}}
\right)
\\
&+& \hat\Lambda^{12} \left(
\frac{9}{64\, a^{10}}
+\frac{55\, \epsilon_1^2}{64\, a^{12}}
+\frac{1495\, \epsilon_1^4}{512\, a^{14}}
+\frac{4035\,\epsilon _1^6}{512\, a^{16}}
\right)
+\mathcal{O}(\hat\Lambda^{16})\nonumber
\\
&=&\frac{1}{2}\epsilon_1 \hat\Lambda
\partial_{\hat\Lambda} \mathcal{W}^{N_f=0, {\rm SU(2)}}.
\end{eqnarray*}

The fact that the eigenvalue of the Mathieu operator is given by\footnote{Precisely, 
the canonical form of the Mathieu equation is 
$\boxed{\psi''(x)+\left(\lambda - 2h^2 \cos 2x \right)\psi(x) = 0}\,$.
Hence, comparing it to eq.~(\ref{M}) one gets 
$$
2x=\varphi,\;\;\;\;
h^2=\frac{4\hat\Lambda^2}{\hbar^2}=\frac{4\hat\Lambda^2}{\epsilon_{1}^{2}},
\;\;\;\;
\lambda=\frac{8\textrm{E}}{\hbar^2}=\frac{8\textrm{E}}{\epsilon_{1}^{2}}.
$$} 
$\hat\Lambda\partial_{\hat\Lambda} \mathcal{W}^{N_f=0, {\rm SU(2)}}$ is an example of the Bethe/gauge 
correspondence discovered by Nekrasov and Shatashvili \cite{NS:2009}.
The Bethe/gauge correspondence 
maps supersymmetric vacua of the ${\cal N}=2$ theories to Bethe
states of some quantum integrable systems (QIS). 
A result of that duality is that
the twisted superpotentials for SU(N) theories are identified with the Yang--Yang (YY) functions
which describe spectra of the corresponding N--particle quantum integrable systems.\footnote{The Yang--Yang
functions are potentials for Bethe equations.}

\begin{figure}[t]\label{Fig1}
\centering
\includegraphics[bb=0 0 1257 942,scale=.3,keepaspectratio=true]{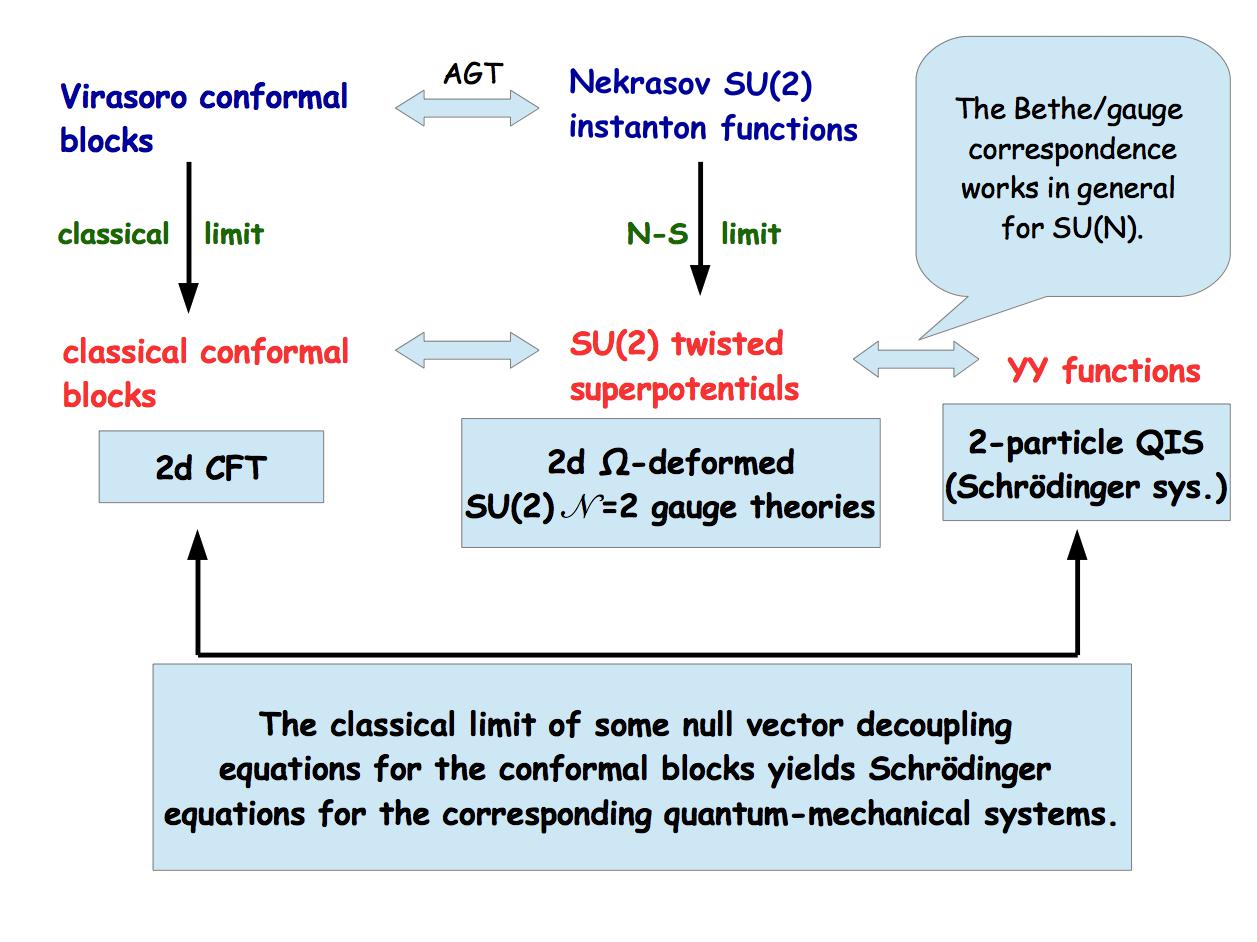}
\caption{The triple correspondence in the case of the Virasoro classical conformal blocks
links the latter to SU(2) instanton twisted superpotentials which describe the spectra of some
quantum--mechanical systems. The Bethe/gauge correspondence on the r.h.s.~connects
the SU(N) ${\cal N}=2$ SYM theories with the N--particle quantum integrable systems. 
An extension of the above triple relation to the case ${\rm N}>2$ needs to consider on the l.h.s.~the 
classical limit of the $W_{\rm N}$ symmetry conformal blocks according to the known extension 
\cite{Wyllard:2009hg} of the AGT conjecture.}
\end{figure}

As has been already mentioned
the twisted superpotentials are defined as the NS limit of the Nekrasov partition functions.
The latter play also a key role in another relation 
--- the celebrated AGT correspondence \cite{Alday:2009aq}. 
Indeed, let $C_{g,n}$ denotes the Riemann surface with genus $g$ and $n$ punctures.
A significant part of the AGT conjecture 
is an exact correspondence between the Virasoro conformal
blocks on $C_{g,n}$ 
and the instanton sectors of the Nekrasov partition functions of certain SU(2) 
quiver gauge theories belonging to some class $S_{g,n}$.
It turns out that the NS limit of the Nekrasov functions corresponds to the so-called classical limit of 
conformal blocks \cite{ZZ}. 
Hence, combining the classical/NS limit of the AGT duality and the Bethe/gauge
correspondence one thus gets the triple correspondence which links the 
classical Virasoro blocks to the SU(2) 
twisted superpotentials and then to spectra of some Schr\"{o}dinger operators (see Fig.\ref{Fig1}).
Indeed, let us note that 2-particle QIS are nothing but the quantum--mechanical systems.

Motivated by the aforementioned correspondences, we have studied in \cite{PP1,pp2,pp3,Piatek:2017fyn} 
examples of a direct link 
between `classical' 2d CFT (i.e., the classical, large central charge
limit of the BPZ equations \cite{BPZ} obeyed by certain {\it degenerate} conformal 
blocks) and the corresponding quantum--mechanical models (= certain stationary 
Schr\"{o}dinger equations), cf.~Fig.\ref{Fig1}.
In sections 2 and 3 of the present article we review main 
results of these investigations. In conclusions we explain 
our motivations and present some open problems for further studies.

In the rest of the introduction, in order to spell out our results, we remind a definition and some 
properties of quantum and classical conformal blocks.
To begin with, let us recall that basic ingredients of any CFT 
model defined on Riemann surface $C_{g,n}$ are 
the correlation functions of the primary fields \cite{BPZ,EO}. Thanks to conformal 
symmetry any correlation function of the primary and descendants fields can be derived once 
the conformal blocks and structure constants are known. The conformal blocks are model independent 
CFT `special functions' defined entirely within a representation theoretic framework.

Indeed, let
${\cal V}_{c,\Delta}^{n}$ denotes the vector space generated by all vectors of
the form:
\begin{equation}\label{basis}
|\,\Delta^{n}_{I}\,\rangle=L_{-I}|\,{\Delta}\,\rangle 
\equiv L_{-k_{1}}\ldots L_{-k_{\ell(I)}}|\,{\Delta}\,\rangle,
\;\;\;\;\;\;\;\;
n=k_1+\ldots+k_{\ell(I)}=:|I|,
\end{equation}
where $I=(k_{1}\geq\ldots\geq k_{\ell(I)}\geq 1)$ is a partition of $n$,\footnote{We 
will use the notation $I\vdash n$.}
$L_n$'s are the Virasoro generators obeying
\begin{equation}\label{Vir}
[L_n, L_m] = (n-m)L_{n+m}+\frac{c}{12}(n^3 - n)\delta_{n+m,0},
\end{equation}
and $|\,\Delta\,\rangle$ is the highest weight state with the following property:
\begin{equation}\label{hw}
L_0 |\,\Delta\,\rangle = \Delta |\,\Delta\,\rangle,
\;\;\;\;\;\;\;\;\;\;\;
L_n |\,\Delta\,\rangle = 0, \;\;\;\;\forall\;n>0.
\end{equation}
The representation of the Virasoro algebra on the space:
$$
{\cal V}_{c,\Delta}=\bigoplus\limits_{n=0}^{\infty}{\cal V}_{c,\Delta}^{n},
\;\;\;\;\;\;\;
{\cal V}_{c,\Delta}^{0} = \mathbb{R} |\,\Delta\,\rangle 
$$
defined by the relations (\ref{Vir}), (\ref{hw}) is called the Verma module
with the central charge $c$ and the highest weight $\Delta$. It is clear that
$\dim{\cal V}_{c,\Delta}^{(n)} = {\sf p}(n)$, where ${\sf p}(n)$ is the number of partitions of 
$n$ (with the convention ${\sf p}(0)=1$).
On ${\cal V}_{c,\Delta}^{(n)}$ exists symmetric
bilinear form $\langle\,\cdot\,|\,\cdot\,\rangle$
uniquely defined by the relations
$\langle\,\Delta\,|\,\Delta\,\rangle=1$
and $(L_n)^{\dagger}\;=\;L_{-n}$.

Let $|\,0\,\rangle$ denotes the vacuum state, i.e.,~the highest 
weight state in the vacuum module with the highest weight $\Delta=0$.
The conformal blocks on the Riemann sphere are defined as the matrix elements  
$$
\langle\,0\,|V_{\Delta_n}(z_n)\ldots V_{\Delta_1}(z_1)|\,0\,\rangle_{\rm sphere}
$$
of compositions of the primary chiral vertex operators (CVO's): 
\begin{eqnarray*}
V_{\Delta_j}(z) &\equiv& V_{\alpha_k,\alpha_i}^{\;\alpha_j}(z):
{\cal V}_{\Delta_i}\To {\cal V}_{\Delta_k},
\;\;\;\;\;\;\;\;
\Delta_{l}=\Delta_{\alpha_l}=\alpha_l(Q-\alpha_l), 
\\
\left[L_n , V_{\Delta}(z)\right] &=& z^{n}\left(z
\frac{\rm d}{{\rm d}z} + (n+1)\Delta
\right)V_{\Delta}(z),\;\;\;\;\;\;\;\;n\in\mathbb{Z}.
\end{eqnarray*}
acting between the Verma modules. The conformal blocks on the torus are traced cylinder matrix elements of
2d Euclidean `space-time' translation operators and CVO's insertions, i.e., 
`chiral partition functions'. The conformal blocks on the higher genus Riemann surfaces can be constructed by 
making use of the so-called {\it sewing} or {\it gluing} procedure.

By inserting projection operators\footnote{$\mathbb{P}_{\tilde\Delta_p}$
are identity operators in ${\cal V}_{c,\tilde\Delta_p}$ built out of the basis vectors (\ref{basis})
and their duals.} 
$\mathbb{P}_{\tilde\Delta_p}$ 
on the intermediate conformal weights $\tilde\Delta_p$, $p=1,\ldots,3g-3+n$
into the internal channels of the 
conformal blocks one gets the latter in terms of the formal power series. 
In the simplest cases, namely, for the 
4-point block on the sphere and the 
1-point block on the torus the coefficients of the power series are implicitly 
defined via recursive relations.
Up to now these coefficient are not been computed in closed form in the general case. 

Taking into account a number of recent applications one of central issues concerning the conformal blocks is 
the existence of their classical limit.\footnote{For a list of some applications, see introduction in 
\cite{Piatek:2017fyn}.}
This is the limit in which all parameters of the conformal blocks
tend to infinity in such a way that their ratios are fixed:
$$
\Delta_i,\; \tilde\Delta_p,\; c\;\To\;\infty,
\;\;\;\;\;\;\;\;\;\;\;\;
\frac{\Delta_i}{c}\;=\;\frac{\tilde\Delta_p}{c}\;=\;{\rm const.}\;\;\;,
$$
$i=1,\ldots,n$, $p=1,\ldots,3g-3+n$. 
For the standard parametrization of the central charge $c=1+6Q^2$, 
where $Q=b+\frac{1}{b}$ and for `heavy' weights 
$(\tilde\Delta_p,\Delta_i)=\frac{1}{b^2}(\tilde\delta_p,\delta_i)$ with 
$\tilde\delta_p,\delta_i={\cal O}(b^0)$
the classical limit corresponds to $b\to 0$.
There exist many convincing arguments, but there is no proof, that in the classical limit the conformal blocks
exponentiate to the functions $f_{\tilde\delta_p}(\delta_i, Z)$ 
known as the classical conformal blocks \cite{ZZ,HJP}, e.g.:
\begin{equation*}
\langle\,0\,|V_{\Delta_n}(z_n)\mathbb{P}_{\tilde\Delta_{n-1}}
\ldots \mathbb{P}_{\tilde\Delta_{2}}V_{\Delta_1}(z_1)|\,0\,\rangle_{\rm sphere}
\;\stackrel{b\to 0}{\sim}
\;{\rm e}^{\frac{1}{b^2}f_{\tilde\delta_p}(\delta_i, Z)},
\;\;\;\;\;\;\;\;\;
Z:=(z_n,\ldots,z_1).
\end{equation*}

The conformal blocks may also include the `light' conformal weights 
${\Delta}_{\sf light}$ which are defined by the property 
$\lim_{b\to 0}b^2{\Delta}_{\sf light}=0$.
It is known, but not proven in general, that light insertions have no influence 
to the classical limit,
i.e.,~do not contribute to the classical blocks:  
\begin{equation}\label{HL}
\left\langle V_{\Delta_{\sf light}}(w)\prod_{i=1}^{n}V_{\Delta_{\sf heavy}}(z_i) \right\rangle_{\!\!\rm sphere}
\;\stackrel{b\to 0}{\sim}\;\Psi(w)\;{\rm e}^{\frac{1}{b^2}f_{\tilde\delta_p}(\delta_i,Z)}.
\end{equation}

Due to the discovery of the AGT correspondence
a considerable progress in the theory of conformal blocks has been recently achieved.
In particular, Gaiotto analyzing an extension of the AGT conjecture to the class of the so-called 
`non-conformal' ${\cal N}=2$, SU(2) super Yang--Mills
theories has postulated the existence of the {\it irregular} conformal blocks \cite{Gaiotto:2009}.
These new types of the conformal blocks were introduced in Gaiotto's work~\cite{Gaiotto:2009}
as products of some new states belonging to the Hilbert space of 2d CFT.
The novel irregular Gaiotto states are kind of coherent vectors for some Virasoro 
generators. It is also known that irregular blocks 
can be obtained from standard (regular) conformal blocks in
properly defined {\it decoupling limits} of the external conformal weights, 
cf.~\cite{Marshakov:2009,Alba:2009fp}. Furthermore, the Gaiotto vectors can be
understood as a result of suitable defined {\it collision limit} of locations of vertex 
operators in their operator product expansion, cf.~\cite{Gaiotto:2012sf}.
Interestingly, also the classical limit makes sense in the case of the irregular blocks.  
This claim first time has clearly appeared in \cite{PP1} as a result of the
non-conformal AGT relations and observations made in \cite{NS:2009}.

\section{Classical limit of irregular blocks and Hill's--type equations} 
\subsection{Hermitian spectral problems}
As mentioned above some 1--dimensional stationary Schr\"{o}dinger equations can be obtained entirely within 
the framework of 2d CFT as the classical limit of the null vector
decoupling (NVD) equations obeyed by certain {\it degenerate} conformal blocks.\footnote{It should 
be emphasized that this claim is consistent with 
the correspondences described above, cf.~Fig.\ref{Fig1}. 
Moreover, let us notice that due to the identification between the classical blocks, the twisted superpotentials 
and Yang-Yang functions, eigenvalues of mentioned Schr\"{o}dinger 
operators can be found by solving appropriate Bethe-like 
equations. Interestingly, very similar in a spirit to what we observe is the so-called ODE/IM correspondence 
\cite{Dorey:1998pt,Bazhanov:1998wj,Dorey:2007zx}. It should be noted here also recently observed link between 
quantum mechanics and topological string theory \cite{Marino}.}
This procedure yields equations some of which are well known in mathematics and physics. For instance, the 
celebrated Mathieu and Whittaker--Hill equations, and some of their solutions
have realizations in 2d CFT. 

Let $|\Delta,\Lambda^2\rangle$ denotes a `pure gauge' or rank $\frac{1}{2}$ 
irregular vector defined by the eqs.:
\begin{gather*}
L_{0}|\Delta,\Lambda^2\rangle =
\left(\Delta + 
\frac{\Lambda}{2} \partial_{\Lambda}\right)|\Delta,\Lambda^2\rangle,
\qquad
L_{1}|\Delta,\Lambda^2\rangle = \Lambda^2 |\Delta,\Lambda^2\rangle,
\\[5pt]
L_{n}|\Delta,\Lambda^2\rangle = 0, \qquad n\geq 2.
\end{gather*}
One can find that the representation of $|\Delta,\Lambda^2\rangle$ 
in terms of the basis vectors (\ref{basis}) in ${\cal V}_{c,\Delta}$ reads as follows:
$$
|\Delta,\Lambda^2\rangle = 
\sum\limits_{n\geq 0}\Lambda^{2n}\sum\limits_{I\vdash n}
\left(G_{c,\Delta}\right)^{(1^n) I}L_{-I}|\Delta\rangle,
$$
where 
$\left( G_{c,\Delta}\right)^{IJ}$ is the
inverse of the \emph{Gram matrix} (or Shapovalov matrix) 
$$
\left(G_{c,\Delta}\right)_{IJ}=\langle\Delta|L_{I}L_{-J}|\Delta\rangle.
$$
The product $\langle\,\Delta,\Lambda^2\,|\,\Delta,\Lambda^2\,\rangle$ 
of pure gauge irregular vectors yields the $N_f=0$ irregular block:
\begin{eqnarray}\label{Nf0}
\langle\,\Delta,\Lambda^2\,|\,\Delta,\Lambda^2\,\rangle =  
\sum\limits_{n\geq 0}\Lambda^{4n} \left(G_{c,\Delta}\right)^{(1^n) (1^n)},
\quad \Lambda = \frac{\hat\Lambda}{\epsilon _1 \epsilon _2}.
\end{eqnarray}
The $N_f=0$ irregular block in the case of the heavy conformal weight $\Delta\sim b^{-2}\delta$ 
and for $\epsilon_2/\epsilon_1=b^2$ exponentiates in the classical limit $b\to 0$
to the classical $N_f=0$ irregular block:
$$
f_{\delta}\!\left(\hat\Lambda/\epsilon_1\right)
= \lim\limits_{b\to 0} b^2 \log\, \langle\,\Delta,\Lambda^2\,|\,\Delta,\Lambda^2\,\rangle
= \sum\limits_{n\geq 1}\left(\hat\Lambda/\epsilon_1\right)^{\!\!4n}\!\!f_{\delta}^{(n)}.
$$
The coefficients $f_{\delta}^{(n)}$ above can be computed order by order from the semi-classical asymptotic
and the expansion (\ref{Nf0}), e.g.:
\begin{gather*}
f_{\delta}^{(1)}= \frac{1}{2 \delta },
\qquad
f_{\delta}^{(2)} = \frac{5 \delta -3}{16 \delta ^3 (4 \delta +3)},
\qquad
f_{\delta}^{(3)} = \frac{9 \delta ^2-19 \delta +6}{48 \delta^5 (4\delta+3)(\delta+2) },
\\[5pt]
f_{\delta}^{(4)} = 
\frac{5876 \delta ^5-16489 \delta
   ^4-22272 \delta ^3+17955 \delta
   ^2+9045 \delta -4050}{512 \delta ^7
   (\delta +2) (4 \delta +3)^3 (4
   \delta +15)} ,
\quad\ldots\;.
\end{gather*}

Let  
\begin{itemize}
\item[(i)]
$\nu$ denotes the Floquet characteristic exponent defined by the property 
$\psi(x+\pi)=\exp(i\pi\nu)\psi(x)$ of the Floquet solution to the Mathieu equation:
\begin{equation}\label{Mathieu}
\psi''(x)+\left[\lambda - 2h^2 \cos 2x \right]\psi(x) = 0\;;
\end{equation}
\item[(ii)]
$\tilde\Delta$ and $\Delta'$ denote the heavy conformal weights related by the fusion rule: 
$$
(\mathrm{I})\quad\tilde\Delta=\Delta\!\left(\sigma-\frac{b}{4}\right), 
\qquad\Delta'=\Delta\!\left(\sigma+\frac{b}{4}\right),
$$ 
where $\Delta(\sigma)=\frac{Q^2}{4}-\sigma^2$;
\item[(iii)]
$V_{+}(z)$ denotes the degenerate primary chiral vertex operator 
with the light degenerate conformal weight:
$$
\Delta_+ :=\Delta_{21}= -\frac{3}{4}b^2-\frac{1}{2}.
$$
\end{itemize}
The non-integer order ($\nu\notin\mathbb{Z}$) Floquet solution of 
the Mathieu equation (\ref{Mathieu}) is \cite{PP1,pp2}:
\begin{eqnarray*}
\psi_{\rm I}(x) &=& {\rm e}^{i\nu x}\,\Psi_{\rm I}(h/2,{\rm e}^{2ix})
= {\rm me}_\nu(x,h)
\\[5pt]
&&\hspace{-40pt}\;=
 {\rm e}^{i\nu x}+\frac{h^2}{4}  \left(\frac{{\rm e}^{(\nu -2)i 
x}}{\nu -1}
-\frac{{\rm e}^{(\nu+2) i x}}{\nu+1}\right)
+\frac{h^4}{32} 
   \left(\frac{{\rm e}^{(\nu +4)
   i x}}{(\nu +1) (\nu
   +2)}+\frac{{\rm e}^{(\nu -4)
   i x}}{(\nu -2) (\nu
   -1)}\right)+\ldots,
\end{eqnarray*}
where
$$
\Psi_{\rm I}(\hat\Lambda/\epsilon_1, z) 
= \lim\limits_{b\to 0} \left(1+\frac{\Phi^{(m\neq n)}_{\rm I}(\Lambda, 
z)}{\Phi^{(m=n)}_{\rm I}(\Lambda)}\right)\,,
\quad z = {\rm e}^{2ix}, 
\quad \frac{\epsilon_2}{\epsilon_1}=b^2,
\quad \Lambda = \frac{\hat\Lambda}{\epsilon _1 \epsilon _2},
\quad h = \pm\frac{2\hat\Lambda}{\epsilon_1}
$$
and
\begin{align*}
\Phi^{(m=n)}_{\rm I}(\Lambda) =&\sum_{n\geq 0}\Lambda^{4n}
\sum_{|I|=n}\left(G^{n}_{c,\Delta'}\right)^{(1^n) I}
\langle\,\Delta'|L_{I}V_{+}(1)L_{-I}|\,\tilde\Delta\rangle
\left(G^{n}_{c,\tilde\Delta}\right)^{I\, (1^n)}\,,
\\
\Phi^{(m\neq n)}_{\rm I}(\Lambda, z) =&
\sum_{\substack{m\neq n\\m,n\geq 0}}
\Lambda^{2(m+n)}z^{m-n}
\sum_{\substack{|I|=m\\|J|=n} }
\left(G^{m}_{c,\Delta'}\right)^{(1^m)\, I}
\langle\,\Delta'|L_{I}V_{+}(1)L_{-J}|\,\tilde\Delta\rangle
\left(G^{n}_{c,\tilde\Delta}\right)^{J\, (1^n)}\,.
\end{align*}
The index I labels first independent solution and means that the fusion rule (I) is assumed.
The corresponding Mathieu eigenvalue $\lambda$
is determined by the $N_f=0$ classical irregular block:
\begin{align*}
\lambda = & \;-\hat\Lambda\,\partial_{\hat\Lambda}\,f_{\delta}
\!\left(\hat\Lambda/\epsilon_1\right)+4\xi^2
\\
=&-\frac{4 h^4}{16}\,f_{\frac{1}{4}-\frac{\nu^2}{4}}^{(1)}-
\frac{8 h^8}{256}\,f_{\frac{1}{4}-\frac{\nu^2}{4}}^{(2)} -
\frac{12 h^{12}}{4096}\,f_{\frac{1}{4}-\frac{\nu^2}{4}}^{(3)} - \ldots + 
4\left(\frac{\nu^2}{4}\right)
\\
=&\;\nu^2 +
\frac{h^4}{2 \left(\nu ^2-1\right)}+
\frac{\left(5 \nu ^2+7\right)h^8}{32 \left(\nu ^2-4\right) \left(\nu 
^2-1\right)^3}+
\frac{\left(9 \nu ^4+58 \nu ^2+29\right)h^{12}}{64 \left(\nu ^2-9\right) 
\left(\nu ^2-4\right) \left(\nu^2-1\right)^5}
+\ldots\;,
\end{align*}
where $\delta=\frac{1}{4}-\xi^2$ and $\xi=\nu/2$.

The above result one gets by considering the classical limit of the 
NVD equation obeyed by the $N_f=0$
degenerate 3-point irregular block 
$\langle\,\Delta', \Lambda^2\,|V_{+}(z)|\tilde\Delta,\Lambda^2\,\rangle$.
A key point in a derivation is the semi-classical asymptotical 
behavior of the type (\ref{HL}), which in the case under consideration takes the 
form:
$$
z^{-\kappa}\,\langle\,\Delta', \Lambda^2\,|\,
V_{+}(z)\,|\,\tilde\Delta,\Lambda^2\,\rangle \;\stackrel{b\to 0}{\sim}\;
\Psi(\hat\Lambda/\epsilon_1, z)\,
\exp\left\lbrace\frac{1}{b^2}f_{\delta}
(\hat\Lambda/\epsilon_1)\right\rbrace,
$$
$\kappa=\Delta'-\Delta_{+}-\tilde\Delta$.
Let us stress that the matrix element 
$\langle\,\Delta', \Lambda^2\,|V_{+}(z)|\tilde\Delta,\Lambda^2\,\rangle$
obeys the NVD equation provided that the fusion rules: (I) or 
$$ 
(\mathrm{II})\quad\tilde\Delta=\Delta\!\left(\sigma+\frac{b}{4}\right),
\quad\Delta'=\Delta\!\left(\sigma-\frac{b}{4}\right)
$$
are assumed. The second possibility determines the second linearly independent solution
$\psi_{\rm II}(x)$, which can be obtained from $\psi_{\rm I}(x)$ by the substitution
$\nu\to-\nu$.

Analogous result holds for the Whittaker--Hill equation \cite{pp3}:
\begin{equation}\label{WH}
\left[-\frac{{\rm d}^2}{{\rm d}x^2}+\frac{1}{2}h^2\cos 4x + 4h\mu\cos 2x\right]
\psi_{\xi}^{\bf 2}
=\lambda^{\mathbf{2}}_{\xi} \,\psi_{\xi}^{\bf 2}.
\end{equation}
Here new objects enter the game, i.e.:
\begin{itemize}
\item the rank $1$ irregular state $|\,\Delta,\Lambda, m\,\rangle$ defined by 
\begin{eqnarray*}
L_{0}|\,\Delta,\Lambda, m\,\rangle
=\left(\Delta+\Lambda\frac{\partial}{\partial\Lambda}\right)|\,\Delta,\Lambda, 
m\,\rangle,
&&
L_{1}|\,\Delta,\Lambda, m\,\rangle  = m\Lambda |\,\Delta,\Lambda,m\,\rangle,
\\
L_{2}|\,\Delta,\Lambda, m\,\rangle =\Lambda^2 |\,\Delta,\Lambda,m\,\rangle,
&&
L_{n}|\,\Delta,\Lambda, m\,\rangle = 0\quad \forall\;n\geq 3
\\[5pt]
&&\hspace{-190pt}\Longleftrightarrow\;\;\;
|\,\Delta,\Lambda, m\,\rangle\;
= \sum\limits_{n\geq0}\Lambda^{n}\sum\limits_{p=0}^{[\frac{n}{2}]} 
m^{n-2p}
\sum\limits_{I\vdash n}\Big(G^{n}_{c,\Delta}\Big)^{(1^{n-2p}\,2^p) I}
L_{-I}|\,\Delta\,\rangle\;;
\end{eqnarray*}
\item the $N_f=2$ irregular block,
$$
\langle\Delta,\tfrac{1}{2}\Lambda,2m_{1}|\Delta,\tfrac{1}{2}\Lambda,2m_{2}\rangle
=\sum\limits_{n\geq 0}\left(\frac{\Lambda}{2}\right)^{2n}\sum\limits_{p,
p'=0 }^{[\frac{n}{2}]}
(2 m_1)^{n-2p}\Big(G^{n}_{c,\Delta}\Big)^{(1^{n-2p}\,2^p)\,(1^{n-2p}\,2^p)}
(2 m_2)^{n-2p'}.
$$
\end{itemize}
Indeed, if we assume the fusion rules (I) or (II) then the equation (\ref{WH}) 
is derived in the classical limit $b\to 0$ from 
the NVD equation obeyed by the 3-point degenerate irregular block,
$$
\langle\Delta',\tfrac{1}{2}\Lambda,2m_{1}|V_{+}(z)|\tilde\Delta,\tfrac{1}{2}\Lambda,2m_{2}\rangle
$$
with 
$$
z={\rm e}^{-2ix},
\quad\quad \frac{h}{2b}=\frac{\hat\Lambda}{\epsilon_1 b}=\Lambda,
\quad\quad m_i = \frac{\hat m_i}{\epsilon_1 b},
\quad\quad \mu=\frac{\hat m_1}{\epsilon_1}=\frac{\hat m_2}{\epsilon_1},
\quad\quad \xi=\nu/2.
$$
The spectrum $\lambda^{\mathbf{2}}_{\xi}$ in eq.~(\ref{WH}) is given in terms of 
the $N_f=2$ classical irregular block
$f_{\delta}^{\bf 2}\!\left(\cdot, \cdot, \cdot\right)$,
i.e.:
\begin{eqnarray*}
\lambda^{\mathbf{2}}_{\xi}\equiv
\lambda^{\mathbf{2}}_{\nu}(h,\mu) 
&=& 
\nu^2-2 h\,\frac{\partial}{\partial h}f_{\frac{1}{4}(1-\nu^2)}^{\bf 
2}\!\left(\tfrac{1}{2}h,\mu,\mu\right)
\\
&=& \nu^{2} + \frac{2}{\nu ^2-1}\mu ^2h^2+\ldots
\;,\quad\quad\quad\quad\nu\notin\mathbb{Z},
\end{eqnarray*}
where in the equation above
\begin{eqnarray*}
f_{\delta}^{\bf 2}\!\left(\hat\Lambda/\epsilon_1, \hat m_1/\epsilon_1, \hat m_2/\epsilon_1\right)
&=& \lim\limits_{b\to 0} b^2 \log\,
\langle\Delta,\tfrac{1}{2}\Lambda,2m_{1}|\Delta,\tfrac{1}{2}\Lambda,2m_{2}\rangle
\\
&=&
\sum\limits_{n\geq 1}
\left(\hat\Lambda/\epsilon_1\right)^{\!2n}
\!f_{\delta}^{{\bf 2},n}\!\left(\frac{\hat m_1}{\epsilon_1},\frac{\hat 
m_2}{\epsilon_1}\right),
\\
f^{\mathbf{2},1}_{\delta}\left(\frac{\hat{m}_1}{\epsilon_1},\frac{\hat{m}_2}{
\epsilon_1}\right) 
&=& 
\frac{1}{2 \delta } \frac{\hat{m}_1}{\epsilon _1} \frac{\hat{m}_2}{\epsilon 
_1},
\\
f^{\mathbf{2},2}_{\delta}\left(\frac{\hat{m}_1}{\epsilon_1},\frac{\hat{m}_2}{
\epsilon_1}\right) 
&=& \frac{\delta ^2 \left(\delta -3 \left(\frac{\hat{m}_2}{\epsilon 
_1}\right){}^2\right)
+\left(\frac{\hat{m}_1}{\epsilon _1}\right){}^2
   \left((5 \delta -3) \left(\frac{\hat{m}_2}{\epsilon _1}\right){}^2-3 \delta 
^2\right)}{16 \delta ^3 (4 \delta +3)},\;\ldots\; .
\end{eqnarray*}
The corresponding non-integer order linearly 
independent solutions $(\psi_{\rm I}^{\bf 2}, \psi_{\rm II}^{\bf 2})$ 
are determined by the fusion rules (I), (II),
and are computable in the same way as in the case of the Mathieu equation.

\subsection{Non--Hermitian PT--symmetric spectral problems with real eigenvalues}
It turns out that analyzing the classical limit of the NVD equations obeyed by
the 3-point degenerate irregular blocks one can get certain solutions
to the eigenvalue problems 
$$
\left[-\frac{{\rm d}^2}{{\rm d}x^2}+Q(x)\right]\psi=\lambda\psi
$$  
with {\it new} complex periodic PT--symmetric\footnote{I.e.~invariant under a
parity (P) reflection and a time (T) reversal.}  
potentials, $\overline{Q(x)}=Q(-x)$, which yield {\it real} spectra $\lambda$. 
The simplest example of such novel class of non-Hermitian and solvable potentials reads as follows:
\begin{equation}\label{generalPot}
Q(x;h,\mu)=\frac{1}{4}h^2{\rm e}^{-4ix}+h\mu\,{\rm e}^{-2ix}+h^2{\rm e}^{2ix}.
\end{equation}
The eigenvalue problem with the potential (\ref{generalPot}) 
is the classical limit of the NVD equation fulfilled by the $N_f=1$ degenerate 
irregular block,
$$
\langle\,\Delta', \frac{1}{2}\Lambda, 2m\,|V_{+}(z)|\,\tilde\Delta, \Lambda^2\,\rangle,
$$
where
$$
z={\rm e}^{-2ix},
\quad\quad\quad\Lambda=\frac{\hat\Lambda}{\epsilon_1 b},
\quad\quad\quad m = \frac{\hat m}{\epsilon_1 b},
\quad\quad\quad\frac{h}{2}=\frac{\hat\Lambda}{\epsilon_1},
\quad\quad\quad\frac{\mu}{2}=\frac{\hat m_1}{\epsilon_1}.
$$
The potential (\ref{generalPot}) is PT-symmetric, 
$\overline{Q(x;h,\mu)}=Q(-x;h,\mu)$, for $x\in\mathbb{R}$
and for real couplings $h,\mu\in\mathbb{R}$. 
In particular, from (\ref{generalPot}) one gets
\begin{itemize}
\item for $\mu=h$ --- the PT--symmetric (for $h^2,x\in\mathbb{R}$) periodic potential discussed in \cite{pp3}:
\begin{eqnarray}\label{Pot1}
Q(x;h,h) &=& \frac{1}{4}h^2{\rm e}^{-4ix}+2h^2\cos 2x 
= \frac{1}{4}h^2\cos 4x + 2h^2\cos 2x - i\frac{1}{4}h^2\sin 4x, 
\end{eqnarray}
\item for $\mu=0$ --- the PT--symmetric (for $h^2,x\in\mathbb{R}$) periodic potential:
\begin{eqnarray}\label{Pot2}
Q\left(x;h,0\right) &=& \frac{1}{4}h^2{\rm e}^{-4ix}+h^2{\rm e}^{2ix}\nonumber
= h^2\left(\frac{1}{4}\cos 4x +\cos 2x\right)+ih^2\left(\sin 2x -\frac{1}{4}\sin 4x\right).
\end{eqnarray}
\end{itemize}
In \cite{pp3} we have computed the non-integer order fundamental 
solutions $(\psi_{\rm I}^{\bf 1}, \psi_{\rm II}^{\bf 1})$
of the Schr\"{o}dinger equation with the potential (\ref{Pot1}). In particular, we have found that
the associated eigenvalue $\lambda:=\lambda_{\nu}^{\mathbf{1}}(h)$
is expressed in terms of the $N_f=1$
classical irregular block $f_{\delta}^{\bf 1}\!\left(\cdot, \cdot\right)$, i.e.:
\begin{eqnarray}\label{h1}
\lambda_{\nu}^{\mathbf{1}}(h) &=& \nu^2-\frac{4 h}{3}\frac{\partial}{\partial h} 
f_{\frac{1}{4}(1-\nu^2)}^{\bf 1}\!\left(\tfrac{1}{2}h,\tfrac{1}{2}h\right)\nonumber
\\
&=&
\nu^2
+\frac{2}{3(\nu^{2} -1) }h^{4} 
+\frac{3}{32 (\nu^{2} -4) (\nu^{2} -1)}h^{6}
+\frac{5 \nu ^2+7}{8 (\nu^{2} -4) (\nu^{2} -1)^3}h^{8}+\ldots\;,
\end{eqnarray}
where $\nu\notin\mathbb{Z}$ and 
\begin{eqnarray*}
f_{\delta}^{\bf 1}\!\left(\hat\Lambda/\epsilon_1, \hat m/\epsilon_1\right)
&=&
\lim\limits_{b\to 0} b^2
\log\,\langle\,\Delta, \frac{1}{2}\Lambda, 2m\,|\Delta, \Lambda^2\,\rangle
=\sum\limits_{n\geq1}
\left(\hat\Lambda/\epsilon_1\right)^{\!3n}
\!f_{\delta}^{{\bf 1},n}\!\left(\frac{\hat m}{\epsilon_1}\right),
\\
f^{\mathbf{1},1}_{\delta}\!\left(\frac{\hat{m}}{\epsilon_1}\right) 
&=& 
\frac{1}{2\delta }\frac{\hat{m}}{\epsilon _1} ,
\\
f^{\mathbf{1},2}_{\delta}\!\left(\frac{\hat{m}}{\epsilon_1}\right) 
&=& 
\frac{5 \delta -3}{16 \delta ^3 (4\delta 
+3)}\left(\frac{\hat{m}}{\epsilon_1}\right)^2
-\frac{3}{16 \delta  (4\delta +3)} ,
\\
f^{\mathbf{1},3}_{\delta}\!\left(\frac{\hat{m}}{\epsilon_1}\right) 
&=& 
\frac{\delta(9 \delta -19)+6}{48 \delta ^5 (\delta+2) (4 \delta 
+3)}\left(\frac{\hat{m}}{\epsilon_1}\right)^3
+\frac{6-7\delta }{48\delta^3 (\delta +2) (4 \delta +3)}\frac{\hat{m}}{\epsilon 
_1},\;\ldots\;.
\end{eqnarray*}
The first few terms in the expansion (\ref{h1}) suggest that the spectrum 
$\lambda_{\nu}^{\mathbf{1}}(h)$
is real for $h^2\in\mathbb{R}$ and $\nu\in\mathbb{R}\setminus\mathbb{Z}$.
Indeed, this observation is true and in elementary way follows from the definition
of the $N_f=1$ irregular block. It should be stressed that an alternative `reality proof' is available here as 
well. Really, one may use the `classical' AGT relation and methods of the dual ${\cal N}=2$ gauge theory employed 
in the calculation of the twisted superpotentials.

\section{Classical limit of regular spherical blocks and the Huen eqution}
In \cite{Piatek:2017fyn} we have continued the line of research described in the previous section, 
this time examining the 2d CFT realization of the Heun equation. Here, one can show that
the classical limit of the second order BPZ NVD equation for the 
simplest two 5-point degenerate spherical blocks: 
$$
\mathscr{F}_{\pm}(z,x)\;:=\;\langle\alpha_4|V_{\alpha_4,\beta_{\pm}}^{\;\alpha_3}(1)
V^{-b/2}_{\beta_{\pm},\beta}(z)
V_{\beta,\alpha_1}^{\;\alpha_2}(x)|\alpha_1\rangle\,,
\;\;\;\;\;\;
\beta_{\pm}=\beta\pm\frac{b}{2},
\;\;\;\;\;\;
\Delta_j=\alpha_j\left(Q-\alpha_j\right)
$$
with $V_{+}(z):=V^{-b/2}_{\beta_{\pm},\beta}(z)$ yields:
\begin{itemize} 
\item[(i)] the normal form of the Heun equation, i.e.,
\begin{equation}
\label{Fuchss2}
\frac{{\rm d}^{2}\Psi}{{\rm d}z^2}+\left[\frac{\delta_1}{z^2}
+\frac{\delta_2}{(z-x)^2} +\frac{\delta_3}{(1-z)^2}
+\frac{\delta_1+\delta_2+\delta_3-\delta_4}{z(1-z)}
+\frac{x(1-x)c_{2}(x)}{z(z-x)(1-z)}\right]\Psi=0
\end{equation}
with the holomorphic accessory parameter $c_{2}(x)$ 
determined by the classical 4-point block on the sphere,
$$
\label{accessory2}
c_{2}(x) =  {\partial\over \partial x}\,
f_{\delta}\!\left[_{\delta_{4}\;\delta_{1}}^{\delta_{3}\;\delta_{2}}\right]\!(x),
$$
where $\left(\Delta, \Delta_i\right)\sim\frac{1}{b^2}\left(\delta, \delta_i\right)$ and
$$ 
{\cal F}_{c,\Delta_\beta}\!\left[^{\Delta_3\,\Delta_2}_{\Delta_4\,\Delta_1}\right]\!(x)
:=\langle\alpha_4|V_{\alpha_4,\beta}^{\alpha_3}(1)
V_{\beta,\alpha_1}^{\alpha_2}(x)|\alpha_1\rangle
\;\stackrel{b\to 0}{\sim}\;
\exp\left\lbrace\frac{1}{b^2}
f_{\delta}\!\left[^{\delta_3\,\delta_2}_{\delta_4\,\delta_1}\right]\!(x)\right\rbrace;
$$
\item[(ii)] the pair of the Floquet--type (path-multiplicative) linearly independent solutions:
\begin{equation}\label{Solutions}
\Psi_{\pm}(z,x)=\lim\limits_{b\to 0}\frac{\mathscr{F}_{\pm}(z,x)}
{{\cal F}_{c,\Delta_\beta}\!\left[^{\Delta_3\,\Delta_2}_{\Delta_4\,\Delta_1}\right]\!(x)}\;.
\end{equation}
\end{itemize}
The first point in the claim above is a well known fact, cf.~e.g.~\cite{LLNZ,FP}.
The second point is an original result of \cite{Piatek:2017fyn}. 
More concretely, in \cite{Piatek:2017fyn} we have derived the formula 
(\ref{Solutions}) and explicitly computed the limit $b\to 0$ by looking into depths of the heavy--light 
factorization of $\mathscr{F}_{\pm}(z,x)$, i.e.:
\begin{eqnarray}\label{Factor}
\mathscr{F}_{\pm}(z,x)
\;\stackrel{b\to 0}{\sim}\;\Psi_{\pm}\left(\infty,1,z,x,0\right)\;{\rm e}^{\frac{1}{b^2}
f_{\delta}\left[^{\delta_3\,\delta_2}_{\delta_4\,\delta_1}\right](x)}.
\end{eqnarray}
In addition we have computed in \cite{Piatek:2017fyn} 
the limit $x\to 0$ of the Heun's solutions (\ref{Solutions})
within 2d CFT framework. 
Work is in progress to answer the question whether this limit solves the trigonometric 
P\"{o}schl--Teller potential.

\section{Discussion}
The results quoted in section 2 are consequences 
of taking the classical limit of the NVD equations obeyed by certain 3-point degenerate irregular blocks defined 
as matrix elements of $V_{+}(z)$ 
between some `simple' (lower-rank) Gaiotto states. One can extend the above analysis to 
the cases of (i) the irregular blocks being matrix elements of $V_{+}(z)$ between the higher-rank
irregular vectors, and (ii) the irregular blocks built out of multi-point insertions of 
$V_{+}(z)\equiv V_{\Delta_{21}}(z)$ or $V_{\Delta_{12}}(z)$ and generic primary CVO's.
It seems that in the first case we should get the Schr\"{o}dinger equation with 
(a) the Hermitian potential built out of higher cosine terms if $N_f = {\sf even}$; (b) 
the non-Hermitian PT--symmetric potential being generalization of (\ref{generalPot}) 
if $N_f = {\sf odd}$.\footnote{Let us notice that the complex periodic
PT--symmetric Hamiltonians with real--valued spectra have fascinating 
applications in the context known as `PT--symmetric complex crystals', cf.~\cite{Bender:2007nj,Longhi}. 
Moreover, the latter idea has amazing experimental realizations in optics, cf.~e.g.~\cite{Longhi}.}

In section 2 we have reported about the weak coupling non-integer order solutions, i.e., the solutions which make 
sense for small couplings $h$, $\mu$ and $\nu\notin\mathbb{Z}$. Therefore, two
interesting questions emerge at this point: (i) how to derive the solutions 
in the other regions of the 
spectra, for instance, for large couplings;\footnote{Note that analogous question has been studied 
for example in \cite{BD} by making use of the $\mathcal{N}=2$ gauge theory tools.} 
(ii) how to extract from the irregular blocks the solutions with integer values of the Floquet 
parameter $\nu$?
We suspect that the answer to the second question hides in the study of the degenerate intermediate conformal 
weight limit of the solutions we have obtained. Our idea of how to tackle the first problem is based on the 
bootstrap technics. We believe, that the relations connecting weak and strong coupling eigenvalue expansions
for Mathieu, Whittaker--Hill and periodic 
PT--symmetric operators are encoded in the classical and decoupling limits
of {\it braiding relation} \cite{HJP2} for the 4-point spherical block.  
Interestingly, this idea is {\it not} hopelessly technically difficult to verify in the case
when one of the external weights is heavy and degenerate.\footnote{Let us note that even in the 
general case of non-rational 2d CFT the classical limit of {\it braiding matrix} is already known, see 
\cite{Verlinde}.}

Finally, using the results quoted in section 3 we plan 
to find 2d CFT realizations of the equations closely related to the Heun equation, i.e.,
the Schr\"{o}dinger equations with the so-called Treibich--Verdier, P\"{o}schl--Teller 
and Lam\'{e} potentials. The latter have also an interesting link with the KdV equation, according to the 
inverse 
scattering method. These investigations should shed some new light on (i) the long-known duality 
between the torus and sphere correlation functions of Liouville theory, and on (ii) the connection problem for 
the Heun equation and its application in black hole physics.



\section*{References}
\begin{thebibliography}{99}
\bibitem{Gorsky:1995zq}
Gorsky A, Krichever I, Marshakov A, Mironov A and Morozov A 1995 
{\it Phys.~Lett.} B {\bf 355} 466-474 
  
\bibitem{Seiberg:1994rs}
Seiberg N, Witten E 1994
{\it Nucl.~Phys.} B {\bf 426} 19-52

\bibitem{Mironov:Morozov:2010}
Mironov A, Morozov A 2010 
{\it J. High Energy Phys.} JHEP04(2010)040

\bibitem{He:2010zzc}
He W 2010
{\it Phys.~Rev.} D {\bf 81} 105017

\bibitem{Maruyoshi:2010}
Maruyoshi K, Taki M 2010 
{\it Nucl.~Phys.} B {\bf 841} 388-425 

\bibitem{NS:2009}
Nekrasov N A, Shatashvili S L 2009 Quantization of Integrable Systems and Four Dimensional Gauge
Theories ({\it Preprint} 0908.4052v1) 

\bibitem{Mironov:2009dv}
Mironov A, Morozov A 2010 
{\it J.~Phys.} A {\bf 43} 195401
  
\bibitem{Nekrasov:Okounkov:2003}
Nekrasov N A, Okounkov A 2003 Seiberg-Witten Theory and Random Partitions  
({\it Preprint} hep-th/0306238v2) 

\bibitem{Nekrasov:2002qd}
Nekrasov N A 2004 
{\it Adv.~Theor.~Math.~Phys.} {\bf 7} 831-864

\bibitem{PP1} 
Piatek M, Pietrykowski A R 2014
{\it J. High Energy Phys.} JHEP12(2014)032

\bibitem{Alday:2009aq}
Alday L F, Gaiotto D and Tachikawa Y 2010 
{\it Lett. Math. Phys.} {\bf 91} 167-197

\bibitem{ZZ}
Zamolodchikov A B, Zamolodchikov A B 1996 
{\it Nucl. Phys.} B {\bf 477} 577-605.

\bibitem{Wyllard:2009hg}
Wyllard N 2009 
{\it J. High Energy Phys.} JHEP11(2009)002

\bibitem{pp2} 
Piatek M, Pietrykowski A R 2016
{\it J. High Energy Phys.} JHEP01(2016)115 

\bibitem{pp3} 
Piatek M, Pietrykowski A R 2016 
{\it J. High Energy Phys.} JHEP07(2016)131

\bibitem{Piatek:2017fyn}
Piatek M, Pietrykowski A R 2017 Solving Heun's equation using conformal blocks 
({\it Preprint} 1708.06135)

\bibitem{BPZ} 
Belavin A, Polyakov A M and Zamolodchikov A 1984 
{\it Nucl. Phys.} B {\bf 241} 333-380

\bibitem{EO}
Eguchi T, Ooguri H 1987
{\it Nucl. Phys.} B {\bf 282} 308-328

\bibitem{HJP}
Hadasz L, Jaskolski Z and Piatek M 2005 
{\it Nucl. Phys.} B {\bf 724} 529-554

\bibitem{Gaiotto:2009}
Gaiotto D 2009 Asymptotically free $N=2$ theories and irregular conformal blocks  
({\it Preprint} 0908.0307)

\bibitem{Marshakov:2009} 
Marshakov A, Mironov A and Morozov A 2009
{\it Phys.~Lett.} B {\bf 682} (2009) 125-129
  
\bibitem{Alba:2009fp}
Alba V, Morozov A 2009 
{\it JETP~Lett.} {\bf 90} 708-712

\bibitem{Gaiotto:2012sf}
Gaiotto D, Teschner J 2012 
{\it J. High Energy Phys.} JHEP12(2012)050

\bibitem{Dorey:1998pt}
Dorey P, Tateo R 1999
{\it J.~Phys.} A {\bf 32} L419-L425

\bibitem{Bazhanov:1998wj}
Bazhanov V V, Lukyanov S L and Zamolodchikov A B 2001 
{\it J.~Statist.~Phys.} {\bf 102} 567-576

\bibitem{Dorey:2007zx}
Dorey P, Dunning C and Tateo R 2007 
{\it J.~Phys.} A {\bf 40} R205 

\bibitem{Marino}
Codesido S, Marino M 2016 Holomorphic Anomaly and Quantum Mechanics 
({\it Preprint} 1612.07687)

\bibitem{LLNZ} 
Litvinov A, Lukyanov S, Nekrasov N and Zamolodchikov A 2014 
{\it J. High Energy Phys.} JHEP07(2014)144

\bibitem{FP}
Ferrari F, Piatek M 2012 
{\it J. High Energy Phys.} JHEP05(2012)025

\bibitem{Bender:2007nj} 
Bender C M 2007
{\it Rept.~Prog.~Phys.} {\bf 70} 947

\bibitem{Longhi}
Longhi S 2011 
{\it J.~Phys.~A:~Math.~Theor.} {\bf 44} 485302

\bibitem{BD}
Basar G,  Dunne G V 2015 
{\it J. High Energy Phys.} JHEP02(2015)160

\bibitem{HJP2}
Hadasz L, Jaskolski Z, Piatek M 2005
{\it Acta Phys.~Polon.} B {\bf 36} 845-864

\bibitem{Verlinde}
Jackson S, McGough L, Verlinde H 2015
{\it Nucl.~Phys.} B {\bf 901} 382-429
\end{thebibliography}
\end{document}